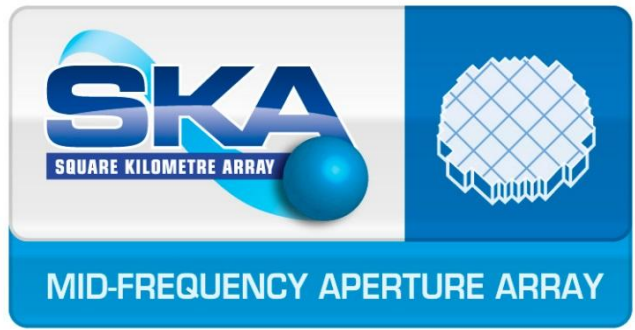

# SKA-AAMID Science Requirements

Document number: SKA-TEL-MFAA-0200009
Context: SE-R
Revision: C
Authors: S.A. Torchinsky/J.W. Broderick/A. Gunst/A.J. Faulkner
Date: 2017-01-23
Document Classification: FOR PROJECT USE ONLY
Status: Released

| Name | Designation | Affiliation | Signature | |
|---|---|---|---|---|
| Authored by: | | | | |
| S. A. Torchinsky | Project Scientist | AAMID | | |
| | | | Date: | 2017-01-23 |
| Owned by: | | | | |
| J. W. Broderick | Project Scientist | AAMID | | |
| | | | Date: | 2017-01-23 |
| Approved by: | | | | |
| Wim van Cappellen | Project lead | AAMID | | |
| | | | Date: | 2017-01-23 |
| Released by: | | | | |
| | | | Date: | 2017-01-23 |

# DOCUMENT HISTORY

| Revision | Date Of Issue | Engineering Change Number | Comments |
|---|---|---|---|
| A | 2015-12-14 | - | First draft by J. Broderick |
|  | 2016-02-25 |  | science motivation, FoV, requirements list, etc |
|  | 2016-03-04 |  | improved text |
|  | 2016-03-15 |  | refining requirements |
|  | 2016-03-25 |  | Refining requirements, and text for science cases |
|  | 2016-03-29 |  | Requirements split into "required" and "goal" |
|  | 2016-04-22 |  | Refining requirement and goals, and science cases |
| 1.19 | 2016-04-29 |  | Version submitted for SRR |
| 1.23 | 2017-01-23 |  | Final version after SRR OAR action items |

# DOCUMENT SOFTWARE

|  | Package | Version | Filename |
|---|---|---|---|
| Wordprocessor | LibreOffice | Oct 2015 | SKA-TEL-MFAA-0200009-SE-SRS-A-SKA-AAMID_Science.odt |
| Wordprocessor | Google docs | Feb 2016 | SKA-TEL-MFAA-0200009-SE-SRS-A-SKA-AAMID_Science |
| Other |  |  |  |

# ORGANISATION DETAILS

| Name | Aperture Array MID Frequency Consortium |
|---|---|
| Registered Address | ASTRON |
|  | Oude Hoogeveensedijk 4 |
|  | 7991 PD Dwingeloo |
|  | The Netherlands |
|  | +31 (0)521 595100 |
| Fax. | +31 (0)521 595101 |
| Website | www.skatelescope.org/mfaa/ |

# Copyright

| Document owner | Aperture Array MID Frequency Consortium |
|---|---|
|  | This document is written for internal use in the SKA project |

# Table of Contents

## LIST OF TABLES

## LIST OF FIGURES

# LIST OF ABBREVIATIONS

| | |
|---|---|
| AAMID | Aperture Array MID Frequency |
| ASTRON | Netherlands Institute for Radio Astronomy |
| AD-n | nth document in the list of Applicable Documents |
| ALIGO | Advanced Laser Interferometer Gravitational-Wave Observatory |
| BAO | Baryon Acoustic Oscillations |
| DM | Dispersion Measure |
| DRM | Design Reference Mission |
| FoV | Field of View |
| FRB | Fast Radio Burst |
| FWHM | Full Width Half Maximum |
| HI | Neutral Hydrogen |
| IM | Intensity Mapping |
| LOFAR | Low Frequency Array |
| LSST | Large Synoptic Survey Telescope |
| MFAA | MID Frequency Aperture Array |
| PSF | Point Spread Function |
| RD-N | nth document in the list of Reference Documents |
| RF | Radio Frequency |
| RM | Rotation Measure |
| RMS | Root Mean Square |
| SETI | Search for Extraterrestrial Intelligence |
| SKA | Square Kilometre Array |
| SKA1 | Square Kilometre Array Phase 1 |
| SKA2 | Square Kilometre Array Phase 2 |
| SKAO | SKA Office |

# 1. Introduction

## 1.1. Purpose of the document

This document describes the top level requirements for the SKA-AAMID telescope as determined by the SKA key science projects. These include parameters such as operating frequency range, instantaneous bandwidth (total processed bandwidth), field of view (or survey speed, as appropriate), sensitivity, dynamic range, polarization purity etc. Moreover, through the definition of a set of science requirements, this document serves as input to a number of other documents contained within this review package (particularly SKA-TEL-MFAA-0200005: 'SKA-AAMID System Requirements' and SKA-TEL-MFAA-0200008: 'MFAA Requirements').

## 1.2. Scope of the document

This document covers the science requirements for the complete SKA-AAMID telescope (Fig. 1). We first provide an overview of the key science cases in Section 3, and then, in Section 4, discuss the various concepts of FoV and how they need to be carefully differentiated for SKA-AAMID. Section 5 contains the list of science requirements, with the salient features of the telescope being summarised in Section 6. Lastly, Section 7 describes a phased approach to the implementation of SKA-AAMID, including possibly early operations in a hybrid combination with the SKA dishes.

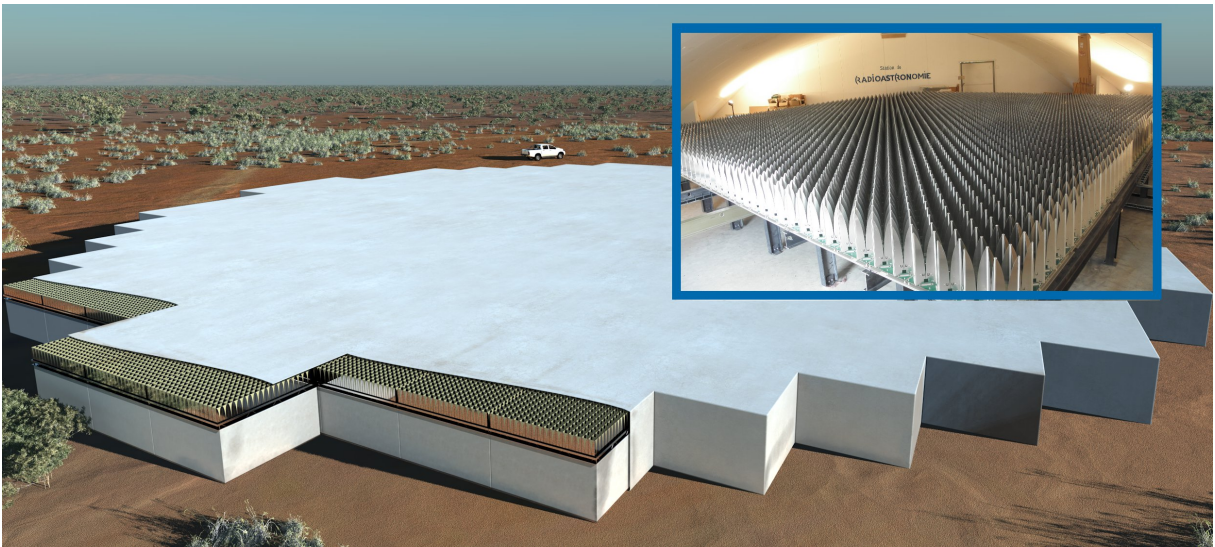

*Figure 1: An artist's conception of an AAMID station. The inset image shows the AAMID pathfinder instrument EMBRACE currently operating at Nancay, France.*

## 2. References

### 2.1. Applicable documents

The following documents are applicable to the extent stated herein. In the event of conflict between the contents of the applicable documents and this document, **the applicable documents** shall take precedence.

| Id | Title | Code | Issue |
|---|---|---|---|
| AD-1 | SKA Request for Proposals | SKA-TEL-SKO-0000020 | 01 |
| AD-2 | Statement of work for the study, prototyping and design on an SKA element | SKA-TEL-SKO-0000021 | 01 |
| AD-3 | Statement of Work for the Study, Prototyping and Preliminary Design of an SKA Advanced Instrumentation Programme Technology | SKA-TEL-SKO-0000022 | 01 |
| AD-4 | SKA Pre construction Top Level WBS | SKA-TEL-SKO-0000023 | 01 |
| AD-5 | SKA System Engineering Management Plan | SKA-TEL-SKO-0000024 | 1 |
| AD-6 | The Square Kilometre Array Intellectual Property Policy | SKA-GOV-0000001 | 1.3 Draft |
| AD-7 | Draft Consortium Agreement | PD/SKA.26-4 | Draft |
| AD-8 | Document Requirements Description | SKA-TEL-SKO-0000029 | 1 |
| AD-9 | SKA Document Management Plan | SKA-TEL-SKO-0000026 | 2 |
| AD-10 | SKA Product Assurance and Safety Plan | SKA-TEL-SKO-0000027 | 1 |
| AD-11 | Change Management Procedure | SKA-TEL-SKO-0000028 | 1 |
| AD-12 | SKA Interface Management Plan | SKA-TEL-SKO-0000025 | 1 |
| AD-13 | SKA1 Level-0 Science Requirements | SKA-TEL-SKO-0000007 | 2 |
| AD-14 | The Radio Frequency Environment At Candidate SKA Sites | WP3-010.020.000-R-001 | 1.2 |

## 2.2. Reference documents

The following documents are referenced in this document. In the event of conflict between the contents of the referenced documents and this document, **this document** shall take precedence.

| Id | Title | Code | Issue |
|---|---|---|---|
| RD-1 | Management Plan | AAMID-TEL.MFAA.MGT-AAMID-MP-001 | 2.0 |
| RD-2 | Delivered Item and Document List | AAMID-TEL.MFAA.MGT.DMT-AAMID-LI-001 | 2.0 |
| RD-3 | The Square Kilometre Array Design Reference Mission: SKA-mid and SKA-lo | DRM | 1.0 |

## 2.3. Reference papers

Abbott B.P. et al., 2016, Phys. Rev. Lett., 116, 061102
Abdalla F.B., Rawlings S., 2005, MNRAS, 360, 27
Abdalla F.B. et al., 2015, Proc. AASKA14, 017
Anderson G.E. et al., 2014, MNRAS, 440, 2059
Ansari R., et al., 2008, arXiv:0807.3614
Arshakian T. G., Beck R., 2011, MNRAS, 418, 2336
Bates S.D. et al., 2013, MNRAS, 431, 1352
Bhat N.D.R. et al., 2004, ApJ, 605, 759
Bignall H. et al., 2015, Proc. AASKA14, 058
Blyth S. et al., 2015, Proc. AASKA14, 128

Braun R. et al., 2009, ApJ, 695, 937
Braun R. et al., 2015, Proc. AASKA14, 174
Brentjens M.A., de Bruyn A.G., 2005, A&A, 441, 1217
Bridle A.H., Schwab F.R., 1999, Synthesis Imaging in Radio Astronomy II, 180, 371
Brown M.L., Battye R.A., 2011, MNRAS, 410, 2057
Brown M.L. et al., 2015, Proc. AASKA14, 023
Bull P. et al., 2015a, Proc. AASKA14, 024
Bull P. et al., 2015b, ApJ, 803, 21
Chu Q. et al., 2016, MNRAS, 459, 121
Condon J.J. et al., 2012, ApJ, 758, 23
Cordes J.M., Lazio T.J.W., 2002, arXiv:astro-ph/0207156
Cordes J.M. et al., 2004, New Astron. Rev., 48, 1459
de Blok E. et al., 2015, Proc. AASKA14, 129
Dewdney P., Landecker T., 1991, IAU Coll. ASP Conf. Series, Vol. 19, pg. 415
Faulkner A. et al., 2010, SKA Memo Series, 122
Fender R.P., Bell M.E., 2011, Bull. Astron. Soc. India, 39, 315
Fender R. et al., 2015a, Proc. AASKA14, 051
Fender R.P. et al., 2015b, MNRAS, 446, L66
Fernández X. et al., 2013, ApJ, 770, L29
Fomalont E.B., 1999, Synthesis Imaging in Radio Astronomy II, 180, 301
Fomalont E., Reid M., 2004, New Astron. Rev., 48, 1473
Gaensler B. et al., 2015, Proc. AASKA14, 103
Hassall T.E. et al., 2013, MNRAS, 436, 371
Haverkorn M. et al., 2015, Proc. AASKA14, 096
Heald G. et al., 2015, Proc. AASKA14, 106
Heidmann J. 1966, L'Astronomie, 80, 157
Heymans C. et al., 2013, MNRAS, 432, 2433
Hopkins A.M., Beacom, J.F., 2006, ApJ, 651, 142
Janssen G. et al., 2015, Proc. AASKA14, 037
Jarvis M. et al., 2015a, Proc. AASKA14, 018
Jarvis M. et al., 2015b, Proc. AASKA14, 068
Johnston-Hollitt et al. M., Proc. AASKA14, 092
Kaplan D.L. et al., 2015, ApJ, 814, L25
Keane E.F., Petroff E., 2015, MNRAS, 447, 2852
Keane E. et al., 2015, Proc. AASKA14, 40
Keane E.F. et al., 2016, Nature, 530, 453
Keith M.J. et al., 2013, MNRAS, 429, 2161
Kloeckner H.R. et al., 2015, Proc. AASKA, 027
Kramer M. et al., 2004, New Astron. Rev., 48, 993
Kulkarni S.R. et al., 2015, arXiv:1511.09137
Laureijs R. et al., 2012, Proc. SPIE, 8442, 84420T
Lee K.J. et al., 2014, MNRAS, 441, 2831
Lorimer D.R. et al. 2006, MNRAS, 372, 777
Lorimer D.R. et al., 2007, Science, 318, 777
LSST Science Collaboration et al., 2009, arXiv:0912.0201
Maartens R. et al., 2015, Proc. AASKA14, 016

McClure-Griffiths N.M. et al., 2015, Proc. AASKA14, 130
Macquart J.-P., Koay J.Y., 2013, ApJ, 776, 125
Macquart J.-P., 2014, PASA, 31, e031
Macquart J.-P. et al., 2015, Proc. AASKA14, 055
McAlpine K. et al., 2015, Proc. AASKA14, 083
McKean J. et al., 2015, Proc. AASKA14, 084
Metzger B.D. et al., 2015, ApJ, 806, 224
Morganti R. et al., 2015, Proc. AASKA14, 134
Mutch S. J. et al., 2011, ApJ, 736, 84
Nakariakov V. et al., 2015, Proc. AASKA14, 169
Norris R. et al. 2015, Proc. AASKA14, 086
Olszewski E.W., et al., 1996, ARA&A, 34, 511
Paragi Z. et al, 2015, Proc. AASKA14, 143
Patel P. et al., 2015, Proc. AASKA14, 030
Peterson J. et al., 2006, arXiv:astro-ph/0606104
Perley R.A., Butler B.J., 2013, ApJS, 204, 19
Petroff E. et al., 2015, MNRAS, 447, 246
Petroff E. et al., 2016, arXiv:1601.03547
Pietka M. et al., 2015, MNRAS, 446, 3687
Popping A. et al., 2015, Proc. AASKA14, 132
Prandoni I., Seymour N., 2015, Proc. AASKA14, 067
Rawlings S. et al., 2004, New Astron. Rev., 48, 1013
Santos M. et al., 2015a, Proc. AASKA14, 019
Santos M. et al., 2015b, Proc. AASKA14, 021
Siemion A. et al., 2015, Proc. AASKA14, 116
Smits R. et al., 2008, SKA Memo Series, 105
Smits R. et al., 2009, A&A, 493, 1161
Smolcic V. et al. 2015, Proc. AASKA14, 069
Sokoloff D.D. et al., 1998, MNRAS, 299, 189
Spitler L.G. et al., 2016, Nature, 531, 202
Staelin D. H., Sutton J. M, 1970, Nature, 226, 69
Staveley-Smith L., Oosterloo T., 2015, Proc. AASKA14, 167
Taylor R. et al., 2015, Proc. AASKA14, 113
Torchinsky S.A., 2009, in Wide Science & Technology for the SKA, p. 33, PoS(SKADS 2009)004
Wilkinson P.N. et al., 2004, New Astron. Rev., 48, 1551
Wilkinson P. 2015, Proc. AASKA14, 065
van der Laan H., 1966, Nature, 211, 1131
van Haarlem M.P. et al., 2013, A&A, 556, A2
Verbiest J.P.W. et al., 2016, MNRAS, 458, 1267
Zhou B. et al., 2014, Phys. Rev. D, 89, 107303

# 3. Scientific motivation

Aperture Arrays are the enabling technology that will lead to the promise of the SKA as a transformational survey machine. The Mid-Frequency Aperture Array (MFAA) is particularly well suited to perform the neutral hydrogen 'billion galaxy survey'. This will enormously advance our understanding of galaxy formation and evolution, and help determine the nature of dark energy. In addition, MFAA will accommodate a range of survey-related science projects profiting from wide field of view, large bandwidth, and the unique multi beaming capabilities. These include all-sky transient monitoring, pulsar searches and bulk timing, as well as wide-field radio continuum surveys, both in total and polarized intensity.

## 3.1. The history of hydrogen in the Universe and the mystery of dark energy

The earliest identified science driver for the SKA is the mapping of HI, including the measurement of the HI mass distribution at cosmological epochs (e.g. Heidmann 1966; Rawlings et al. 2004), and mapping HI in the Galaxy (e.g. Landecker & Dewdney 1991). Indeed, an all-sky 'billion galaxy survey', using the redshifted HI 21-cm line, is at the heart of the SKA2 science case (e.g. Abdalla & Rawlings 2005; Staveley-Smith & Oosterloo 2015). SKA-AAMID, with its wide field of view and exquisite sensitivity, is the ideal survey instrument to obtain this unique, high-legacy source of information.

By tracing the HI of individual galaxies over the entire sky and at different redshifts, SKA will be able to detect baryon acoustic oscillations (BAO), which constitute a preferred clustering scale in the distribution of matter on cosmological scales, and its evolution as a function of cosmic time (e.g. Abdalla et al. 2015; Bull et al. 2015a). This will enable a direct measurement of the effects of dark energy (e.g. Rawlings et al. 2004; Abdalla & Rawlings 2005); in particular, below a redshift of 2, dark energy will start to dominate the Universe's expansion, and BAO detections will enable important distinctions between different cosmological models. The optimal redshift range for this study is between $z = 0.5 - 2$, corresponding to a frequency range of 450 to 1000 MHz. A large processed FoV of order hundreds of square degrees will be critical in accessing large numbers of galaxies within a few years of observing time. Angular resolution on the order of 5 arc seconds is needed to beat the confusion limit. With these data, SKA-MFAA will provide constraints on the Equation of State of the Universe, and may well indicate that Einstein's theory of General Relativity, together with a Cosmological Constant, is insufficient to explain dark energy.

The method of intensity mapping (IM) has been proposed to make a statistical detection of the BAO in the HI mass distribution (e.g. Peterson et al. 2006; Ansari et al. 2008; Santos et al. 2015a; Bull et al. 2015b). It is likely to detect the first peak in the power spectrum, and a modest-sized SKA-AAMID initial station can achieve this result. Moreover, HI emission observations are not affected by, for example, dust extinction, which may complicate competing optical/NIR projects such as the *Euclid* space mission (Laureijs 2012). Only the full SKA-AAMID can make a detailed map of the HI distribution in an unprecedented cosmic volume. It will enable measurements of the higher-order features in the power spectrum, which are necessary to properly match the signature found in the Cosmic Microwave Background (Abdalla & Rawlings 2005).

## 3.2. Galactic structure and star formation

Star formation is a complex process involving interactions on length scales spanning many orders of magnitude. The study of star formation in the Milky Way requires observations of regions as small as the Solar system, while at the same time, the matter fuelling the star formation is flowing in and out on Galactic scales. As a result, it is necessary to understand the structure in this flow of material, including the large scale of the 'Galactic fountain' and throughout the many length scales of turbulence down to Solar system sized over-densities where the stars are forming (e.g. McClure-Griffiths et al. 2015).

Neutral hydrogen is the primary material for star formation, arriving from the intergalactic medium, as well as from nearby galaxies such as the Magellanic System. SKA-AAMID is ideally suited to trace the HI at angular scales spanning many orders of magnitude, with short baselines within stations and out to long baselines on the order of 100 km.

Neutral hydrogen can also be measured in absorption by observing along lines of sight towards distant strong continuum sources, such as quasars. In this way, we probe cool gas, which has formed pockets after splitting to form the Cool Neutral Medium, or fast moving High Velocity Clouds. This method also allows us to measure the local spin temperature, which is a crucial parameter for understanding the dynamics leading to star formation. With SKA-AAMID, we will have access to many sources in the large FoV to probe the Galaxy for cool gas in multiple directions.

The Milky Way is soon entering a phase of evolution seen in other galaxies called the 'Green Valley' during which there is no longer star formation (Mutch et al. 2011). It remains unclear why star formation is quenched, and in-depth studies of the Milky Way will teach us more about this important phase of evolution, and the role it plays in galaxy evolution.

The Magellanic System, including the Large and Small Magellanic Clouds (LMC and SMC), provides material to the Milky Way, but it is also the site of star formation (Olszewski et al. 1996). Studies of the Magellanic clouds provides us with a nearby laboratory for observing star formation in a low metallicity environment, giving us clues about the earliest star formation in the history of the Universe. For these studies, the telescope must be in the southern hemisphere in order to observe the LMC and SMC, which are at a declination of approximately -70˚. The Karoo site is far enough south to observe the LMC/SMC, but they reach a maximum elevation of only 48˚ so observing time is somewhat limited. This could be improved by having access to the sky at elevations below 45˚.

## 3.3. The structure of the Universe

A radio continuum survey comprising over three billion galaxies will also be carried out with SKA-AAMID (an order of magnitude more than with SKA1; e.g. Prandoni & Seymour 2015; Norris et al. 2015). When combined with the HI survey discussed in Section 3.1, this survey will ensure major advances in state-of-the-art precision cosmology. By probing the distribution and evolution of large-scale structure over ultra-large volumes, key parameters associated with dark matter, dark energy, modified gravity and primordial non-gaussianity will be able to be determined to an accuracy well beyond that which is possible with SKA1. Moreover, SKA2 will surpass optical/near-infrared projects such as *Euclid* and the Large Synoptic Survey Telescope (LSST; LSST Science Collaboration et al. 2009), although there will be valuable and important synergies by combining data from several facilities (e.g. Maartens et al. 2015). SKA-AAMID alone can conduct a unique test of the cosmological principle of isotropy by confirming if the direction of the matter dipole matches that of the dipole evident in the Cosmic Microwave Background (e.g. Maartens et al. 2015; Jarvis et al. 2015a).  Owing to the expected significant improvements in sensitivity and survey speed of SKA-AAMID over SKA1-MID, it may even be possible to directly measure the expansion rate of the Universe in real time, by detecting very small frequency shifts in the HI signal of galaxies over the expected 50-year lifetime of the SKA-AAMID facility (Kloeckner et al. 2015).

The effects of weak gravitational lensing of distant galaxies is an important part of the SKA2 science case; this can be investigated by measuring the morphological curvature in the SKA-AAMID large sample of galaxies. This in turn gives a measure of the dark matter in the Universe, and, in general, a probe of late time cosmology (e.g. Heymans et al. 2013). The analysis will not only complement optical surveys having completely different systematic error effects, but the radio survey has a number of important advantages which will surpass optical surveys of cosmic shear (Brown et al. 2015).  The combination of the radio continuum survey with the HI spectroscopic survey will allow a clear separation of populations between the lens galaxies and the lensed galaxies, thus reducing a source of systematic error.  The spectroscopic survey will also provide rotation curves of lensed galaxies which can be compared to the morphological major axis of the galaxy.  In the absence of lensing, the galaxy major axis and the axis of rotation are orthogonal, so a departure from orthogonality is an independent measure of gravitational lensing which can be used to reduce systematic error in large samples of lensed objects.  Polarization can also be used as a complementary probe since polarization angle is not affected by gravitational lensing (Brown & Battye 2011). Detailed polarization maps of lensed objects studies permits a determination of the original unlensed morphology of the object which is effectively a direct measure of the mass of the lens object.

An SKA-AAMID continuum survey will detect almost all star-forming galaxies that are seen in the optical with LSST (Norris et al. 2015).  The HI survey will be highly complementary: optical-quality imaging (i.e. arcsec resolution) of galaxies the size of our Milky Way will become routine out to a redshift of 1, when the Universe was about 40 per cent of its current age (e.g. Blyth et al. 2015). More locally, highly-detailed HI studies of the interstellar medium will be possible for hundreds of nearby galaxies (e.g. de Blok et al. 2015), at a resolution currently only possible for just a handful of very close objects (e.g. Braun et al. 2009). Furthermore, on larger scales, huge reservoirs of gas in the 'cosmic web' will be able to be seen in unprecedented detail (e.g. Popping et al. 2015). By combining these features, the life cycle of gas in galaxies, particularly how gas initially enters galaxies to enable star formation to take place, as well as the interplay between the star formation rate and black hole growth/accretion, can be explored across nearly 13 billion years of cosmic time (e.g. Jarvis et al. 2015b; Smolcic et al. 2015; McAlpine et al. 2015). This is particularly important at $z = 2$, where the star formation rate was at its peak, over an order of magnitude higher than in the present-day Universe (e.g. Hopkins & Beacom 2006). SKA-AAMID has the frequency coverage, angular resolution and sensitivity to study this crucial epoch for the first time in both continuum and HI.

## 3.4. The dynamic radio sky

### 3.4.1. Transients and variables

Our understanding of the dynamic radio sky will be revolutionized with SKA-AAMID. The implementation of mid-frequency aperture array technology will result in enormous, multiple FoVs; these will be combined with unprecedented sensitivity, high time resolution, as well as the precise localization accuracy offered by long baselines. The near-real-time reduction of the resulting continuous, commensal, all-sky data streams, as well as extremely rapid responses to external triggers from other facilities, will allow a highly-detailed census of luminous cosmic explosions: the sites of extremes in density, pressure, temperature and gravity, far beyond what is possible in a terrestrial laboratory (e.g. reviews by Fender & Bell 2011; Fender et al. 2015a). The SKA will then truly be a radio sky monitor, shedding light on a rich, virtually unexplored parameter space spanning over 25 orders of magnitude in luminosity, and over time-scales from nanoseconds to years (e.g. Cordes et al. 2004; Pietka et al. 2015).

It is crucial to note that the key metric for transient studies is detection rate, and that this is linearly proportional to FoV (e.g. Macquart 2014). Thus, aperture arrays are essential for making transformational advances in this science area. While the low-frequency aperture array will already be deployed in SKA1, a mid-frequency aperture array will operate in a frequency regime that is (i) less affected by scattering (standard frequency dependence $v^{-4}$; e.g. Bhat et al. 2004); (ii) in which the sky temperature is lower, resulting in better instantaneous sensitivity; and (iii) where the angular resolution is finer for a given maximum baseline, thus allowing more precise localization of a transient event. Hence, SKA-AAMID will be the facility of choice for surveying the transient and variable radio sky, with a particular 'sweet spot' for bright, coherent emission from 'fast' transients, including the most rare, extreme outbursts in the Universe.

Another critical advantage of aperture arrays for transient science is that they are 'software telescopes' with no moving parts. Hence they have the ability to react within seconds to external robotic triggers, or internal triggers from the SKA itself. It is this capability for rapid response that is absolutely essential for high-impact, cutting-edge science (e.g. Anderson et al. 2014; Fender et al. 2015b; Petroff et al. 2015; Kaplan et al. 2015). On the other hand, with dish arrays, the technique of sub-arraying will be required to cover large parts of the sky at any moment, carrying a significant sensitivity penalty. This is not nearly as efficient as the aperture array solution.

The discovery of 10,000 or more 'fast radio bursts' (FRBs; e.g. Lorimer et al. 2007, Thornton et. al. 2013) at cosmological distances will be a unique probe of the dark energy equation of state, as well as the magnetic fields in the intervening intergalactic medium (e.g. Macquart & Koay 2013; Zhou et al. 2014; Macquart et al. 2015). At the time of writing, 17 FRBs had been discovered, all but one at 1400 MHz, with the remaining event being detected at 800 MHz (Petroff et al. 2016). Clearly, much will be learned about these enigmatic sources with the current generation of radio facilities (e.g. recent results by Keane et al. 2016; Spitler et al. 2016), as well as SKA1. However, it is interesting to note that SKA-AAMID will cover the frequencies where FRBs have already been detected, and that, at least in some cases, FRBs perhaps may not be able to be detected as easily with low-frequency aperture arrays (Kulkarni et al. 2015). Hence, with a field of view of at least several hundred square degrees at 1000 MHz, SKA-AAMID is in a very strong position to be the prime facility for discovering FRBs in sufficient numbers to do world-class cosmology.

In addition, SKA-AAMID will be very sensitive to more slowly varying, incoherent synchrotron transients, which will be detected at rates of up to 1000 per week (Burlon et al. 2015; Fender et al. 2015a), allowing us to move from current standard individual source studies to detailed population statistics. While one can argue that frequencies of a few GHz and above offer the best prospects of detections (assuming, for example, a van der Laan 1966 model for a given outburst), commensal and dedicated SKA-AAMID surveys around 1000 MHz will still be vital in transforming our knowledge of the complex zoo of slow transients (e.g. forecasts by Metzger et al. 2015), as well as variability statistics in general (e.g. as discussed by Bignall et al. 2015 in the context of AGN and extreme scattering events). As just one example, an all-sky survey with an rms noise level of a few µJy beam$^{-1}$ and at arcsecond resolution could be performed on a daily basis, helping to open up entirely new parameter space, as discussed above.

### 3.4.2. Pulsar astronomy

Pulsar astronomy with SKA2 will have a profound impact on our understanding of the fundamental tenets of General Relativity. SKA-AAMID will be able to find the vast majority of the expected 40,000 visible pulsars in our Galaxy, enabling more than a ten-fold increase on the known pulsar population of today (e.g. Smits et al. 2009; Keane et al. 2015). A definitive census of the complex zoo of neutron stars will then be possible, including new discoveries at the extremes of parameter space such as the 'holy grail' pulsar-black hole binary systems. Such systems may be so rare that there are only a few existing in the entire Milky Way; the only way to find them is to detect close to all the active pulsars in the Galaxy. Studies of the equation of state of ultra-dense matter, and strong-field tests of relativistic gravity, will benefit substantially from this new paradigm. For example, completely novel tests of the basic characteristics of a black hole – mass, charge and spin – as well as the presence of an event horizon, will be possible (e.g. Kramer et al. 2004).

The wide field of view, large instantaneous bandwidth, and multi-beaming capabilities make SKA-AAMID an unrivalled pulsar finding and timing machine. Excluding a few degrees on either side of the Galactic Plane, pulsar surveys are most effective over the approximate frequency range 550 - 900 MHz (Smits et al. 2008,2009). Surveys near and along the Galactic plane can be carried out at slightly higher frequencies: approximately 1100-1400 MHz. A large field of view makes the survey time efficient, while the large bandwidth enables an accurate estimate of dispersion measure (DM) as well as the potential effects of scatter broadening. The multi-beaming allows highly efficient monitoring and timing, while eliminating instrumental effects in the calibration by observing multiple pulsars simultaneously. Moreover, a southern location for the telescope is ideal for surveying along the Galactic plane.

An exciting new era of astrophysics has recently begun, with the direct detection of gravitational waves with the Advanced Laser Interferometer Gravitational-Wave Observatory (ALIGO; Abbott et al. 2016). SKA2 will also be a gravitational wave detector, using an ensemble of ultra-stable, isolated millisecond pulsars, well distributed on the sky (Janssen et al. 2015; also see e.g. Verbiest et al. 2016).

Long-term, very-high-precision observations of such a pulsar timing array will allow the detection of long-period gravitational waves from, for example, galaxy collisions and the subsequent mergers of their central supermassive black holes, or from more exotic phenomena such as cosmic strings. This will allow further crucial tests of gravity in the strong-field regime, and new constraints for models of galaxy evolution. Moreover, the nHz frequencies of these gravitational waves will complement the much higher frequencies accessible to facilities such as ALIGO.

At present, insufficient suitable candidates for the timing array are known, and the dedicated pulsar survey described above is essential to establish a dense network of new candidates. These sources then need to be monitored for an extended period (e.g. once every two weeks for about 6 months; Smits et al. 2009) to determine which are most suitable for high precision timing, which is best done at frequencies of at least a few GHz (Janssen et al. 2015). However, SKA-AAMID is not only ideal for finding pulsars, but also for the initial 'bulk timing' consisting of multiple observations of new discoveries to obtain basic parameters: spin period, binary parameters, spin down etc. It will also be a vital facility for tracking dispersion measure variations in the high-precision timing experiment, owing to its large bandwidth, and there will be completely novel pulsar timing experiments only possible with SKA-AAMID (e.g. semi-continuous timing using dedicated beams).

## 3.5. The origin and evolution of cosmic magnetism

The often overlooked contribution of cosmic magnetic fields is beginning to be ameliorated. This is due both to the advent of new analysis techniques (e.g. Faraday rotation measure (RM) synthesis; Brentjens & de Bruyn 2005), as well as telescopes with improved sensitivity and large instantaneous bandwidths of a 1000 MHz or more. Nonetheless, SKA2 will be required to enter an era of very-high-precision magnetism science, in which the origin and evolution of cosmic magnetic fields can be fully explored from the distant to local Universe.

SKA-AAMID has a number of features that make it very attractive for magnetism studies. The most important is a wide bandwidth at and below approximately 1000 MHz: this ensures excellent resolution in Faraday depth (i.e. RM) space, where such a resolution is inversely proportional to the $\lambda^2$ baseline covered by the telescope. An AAMID telescope operating between 450 – 1450 MHz has the requisite Faraday depth resolution, sufficient sensitivity to extended Faraday depth structures, as well as the ability to detect very large absolute Faraday depths, making it an excellent telescope to study a wide range of magnetised phenomena, both extragalactic and within the Milky Way (e.g. Gaensler et al. 2015; Haverkorn et al. 2015). Apart from increasing the $\lambda^2$ coverage, the lower operating frequencies would allow progressively higher-redshift sources to be targeted (e.g. Arshakian & Beck 2011; Gaensler et al. 2015). SKA-AAMID also has sufficient angular resolution to reduce the deleterious effects of beam depolarization.

Key SKA2 projects such as a dense, all-sky RM grid of at least 40 million radio sources (Johnston-Hollitt et al. 2015), high-resolution Faraday tomography of nearby galaxies (Heald et al. 2015), and studies of the faint polarized sky at the level of a few µJy or less (Taylor et al. 2015) all require survey speed that only SKA-AAMID can deliver. In addition, mid frequencies offer an ideal compromise between the usually competing effects of continuum brightness level (assuming a canonical spectral index α = -0.7) and internal Faraday depolarization (e.g. Sokoloff 1998).

## 3.6. Additional science

Many science cases for the SKA will benefit from the unique capabilities of mid-frequency aperture array technology. While we have described a number of key science areas above, an additional example that would utilize the very wide FoV is commensal, ultra-sensitive searches for extraterrestrial intelligence (SETI; e.g. Siemion et al. 2015). More locally, Solar imaging in the frequency range 450 – 1000 MHz is currently poorly explored, and SKA-AAMID will have the combination of baseline coverage, field of view and modes of operation that would allow it to be significantly more flexible for such a task than the SKA1 dishes (Nakariakov et al. 2015).

Furthermore, through the exploration of previously uncharted parameter space, SKA-AAMID will undoubtedly also result in many new discoveries (e.g. Wilkinson et al. 2004; Wilkinson 2015). In fact, most telescopes have made unexpected discoveries completely outside the domain for which the instrument was designed (e.g. review by Torchinsky 2009). The 'unknown' covers not just time-domain science: for example, the combination of multi-beaming, along with unprecedented instantaneous FoV, enables detailed studies of structures that span tens of degrees in angular size on the sky. Thus, from our own Galaxy out to cosmological distances, SKA-AAMID has the immense potential to transform our knowledge of the Universe, while fully realizing the philosophy of the SKA as a scientific facility for exploring the unknown.

## 4. Definition of 'Field of View'

The term 'Field of View' (FoV) is used throughout SKA documentation, but the definition is not always clear.

There are two definitions, which can be qualified by 'optical' and 'processed'.

## 4.1. Optical Field of View

The optical field of view is the maximal solid angle on the sky in which beams can be formed instantaneously. Since the processed field of view (see below) is not necessarily contiguous, the optical field of view is mostly larger than, or equal to, the processed field of view. It is determined by the electromagnetic response of the frontend receptor. For dishes, the optical FoV is a function of the ratio of focal length to dish diameter. This determines how far from the centre of the focal spot one can place a detector before there is unacceptable aberration. For aperture arrays, the optical FoV is determined by the antenna response (point spread function), and also the maximum acceptable scan angle (angle measured from zenith). In general, the FWHM of the smallest antenna element determines the optical FoV. If we group antenna elements together either by analogue or digital summing, then this grouping becomes the smallest antenna element, and the optical FoV is limited by the FWHM of this grouping. It is possible for an aperture array FoV to be constrained by the maximum scan angle, rather than the FWHM of the antenna element. The optical FoV is frequency dependent.

## 4.2. Processed Field of View

The processed field of view is the sum of all solid angles on the sky that can be instantaneously observed and processed. It is determined by the digital processing capacity of the backend (correlator or digital beamformer). There is a processing bandwidth, which can be shared between the number of beams formed and the bandwidth of each beam. In general, it is the processed FoV which puts the limitation on FoV, and this is the one of interest. However, it is possible to have a processed FoV which exceeds the optical FoV by forming beams outside the optical field of view. In these cases, performance will be compromised. For example, for a LOFAR station, one can form a maximum of 488 digital beams, each with a bandwidth of only 200 kHz (van Haarlem et al. 2013). However, tiling the sky with this many beams would put many of them well below the 45° scan angle limit, and those beams would have severely compromised performance because of the optical limitation of scan angle and antenna response.

## 4.3. Use of the term FoV in science documentation

The term FoV is used throughout SKA science documentation, and in general, this refers to the processed FoV. However, there may sometimes be confusion. For example, in SKA-AAMID documentation, one advantage which is often mentioned is the ability to "form multiple fields of view". In fact, in that case, the 'field of view' is determined by the PSF of the analogue summed antenna elements. We split the signal and perform analogue beamforming on the copies, and continue with digital processing of the multiple copies of the analogue signal. This is effectively a method to increase the processed FoV. One might ask the question, why do analogue summing, only to split the signal again and digitize? We do this to achieve a processed FoV which is larger, but not as large as what would be achieved by digitizing the antenna elements. In this way, we have a larger processed FoV, but with fewer digitizers than required to process the entire optical FoV which would require the digitization of each individual antenna element. If we have sufficient "multiple fields of view" to cover the entire optical FoV, then there is no sense in making multiple fields of view, and we should just digitize the individual elements.

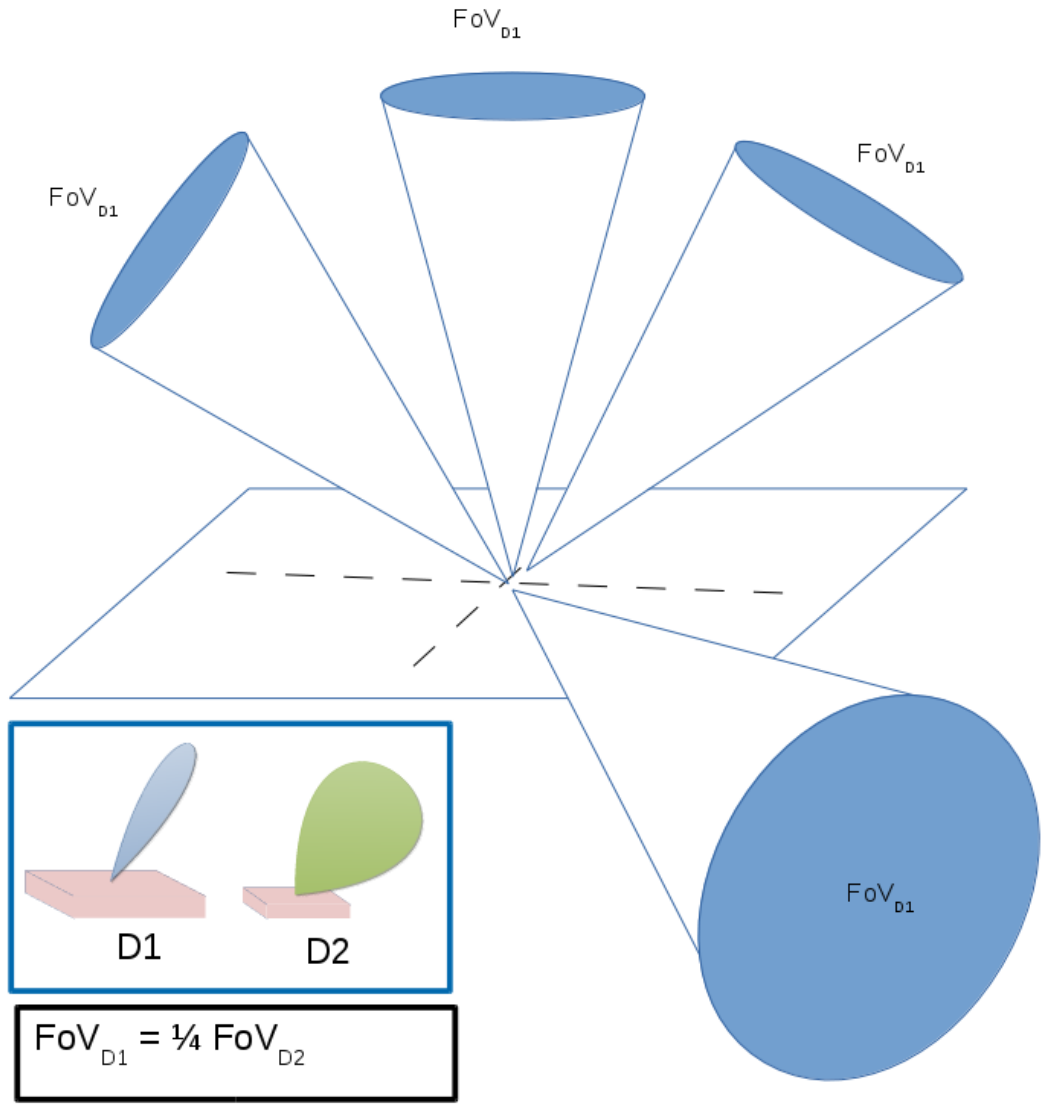

*Figure 2: A schematic illustrating the concept of multiple fields of view. In this illustration, Design-2 (D2) has a small tile as the basic element. The antennas on the tile are summed together with analogue beamformer circuits applying either phase shift or time delay. Design-1 (D1) takes the analogue beamforming a further step, and is a grouping of 4 tiles. As a result, the beamsize of D1 is smaller than D2 by a factor of 4 in terms of angular area. Design-2 therefore has 4 times the field of view of D1. This field of view can be recovered in D1 by by forming 4 independent output beams from the D1 group of tiles, which are then treated independently. As a result, 4 beams are created which can be pointed in arbitrary directions.*

## 4.4. The FoV advantage of SKA-AAMID, and future upgrades

The main advantage of SKA-AAMID is its large optical FoV. While it may be true that we do not currently have the capacity to produce a processed FoV equal to the optical FoV, the SKA is expected to operate for 50 years and more. SKA-AAMID provides the opportunity for future upgrades to increase the processed FoV without changing the frontend hardware. This is not the case for dish-based solutions which will be limited by a much smaller optical FoV compared to SKA-AAMID.

## 5. Science Requirements

The full SKA telescope is envisaged to be a facility approximately one order of magnitude more sensitive than SKA1, with additional significant upgrades to the FoV (at least 20 times larger) and maximum baseline (as large as 3000 km). We now describe a range of preliminary science requirements for an AAMID component in SKA2. It is important to stress that the final requirements will naturally evolve depending on the scientific outcomes and lessons learned from SKA1. Nonetheless, the values described henceforth allow important initial constraints to be placed on the overall design of SKA-AAMID.

To derive the requirements, we have used references in the literature, chapters from the SKA Science Book 'Advancing Astrophysics with the Square Kilometre Array' (summary chapter Braun et al. 2015), as well as the SKA Design Reference Mission (DRM, document RD-3) and the latest version of the SKA1 Level 0 science requirements (document AD-13).

Note that henceforth, the term 'beamformed' refers to high-time-resolution, pulsar-like observing modes. Multiple beams may be formed, and each beam is the result of either the coherent ('tied-array') or incoherent summation of the output of all SKA-AAMID stations, each producing high-time-resolution output. Furthermore, the term 'imaging' refers to interferometric observations, where a number of beams are used to tile a sky region of interest.

In addition, although the following specifications are stated for a number of particular durations, this is not a limitation on observation length, but merely a time-scale for a given performance. Moreover, unless otherwise stated, the requirements are necessary within the entire MFAA tuning range of 450 to 1450 MHz. Finally, HI requirements often use an upper limit of 1425 MHz, to account for the possibility of nearby blueshifted HI.

### 5.1. General considerations

**AMSC_R1001**:

*SKA-AAMID shall be sited in the Karoo Region of South Africa, at a specific location as decided by the SKA Members.*

This location, the site of SKA1-MID, is ideal for the full range of Galactic and extragalactic science cases. It will allow sharing of infrastructure already on the ground during SKA1, as well as the possible hybrid combination of SKA-AAMID and the SKA dishes (Section 6). Moreover, the telescope will be at a similar latitude to a number of current premier optical/NIR facilities, as well as the planned next generation of these telescopes, including LSST, the European Extremely Large Telescope (E-ELT), and the Giant Magellan Telescope (GMT). The latitude of the Karoo also allows the telescope access to a number of current and planned deep fields with multi-wavelength ancillary data (e.g. RD-3, Prandoni & Seymour 2015). Lastly, the near-pristine radio-frequency interference (RFI) environment (e.g. AD-14) is also vital for numerous scientific objectives.

**AMSC_R1002:**

*SKA-AAMID shall provide continuous frequency coverage of both perpendicular polarization states from 450 - 1450 MHz. It is desirable to extend the frequency coverage to the (continuous) range 400 - 1500 MHz.*

This fundamental requirement is mandatory to ensure that cutting edge, mid-frequency science can be carried out with the telescope.

**AMSC_R1003a:**

*SKA-AAMID shall have an instantaneous, possibly non-contiguous, processed FoV ≥ 100 deg$^2$ at 1000 MHz, where this processed FoV is synthesized from multiple beams. A $v^{-2}$ frequency profile for the FoV is assumed for the other operating frequencies.*

**AMSC_R1003b:**

*The (possibly non-contiguous) instantaneous bandwidth per beam shall be ≥ 500 MHz.*

**AMSC_R1003c:**

*The product of processed FoV and bandwidth shall be constant, allowing larger FoVs to be realised by using smaller bandwidths.*

The flexibility of choosing observing configurations for SKA-AAMID ties together FoV, instantaneous bandwidth, and the number of beams. Apart from detections of transients, a wide FoV is vital for sufficiently high survey speeds for a range of other science topics. Wide instantaneous bandwidths are also important for a number of reasons, namely sensitivity, survey speed (e.g. covering a large redshift range in extragalactic HI studies), spectral index determination (particularly for time-variable phenomena), interference rejection, coherent de-dispersion and scatter-broadening measurements for pulsars, as well as high Faraday depth resolution using the RM synthesis technique.

**AMSC_R1004a:**

*The total optical FoV shall be ≥ 200 deg$^2$ at 1000 MHz, with a $v^{-2}$ frequency profile for the FoV being assumed for the other operating frequencies.*

**AMSC_R1004b:**

*SKA-AAMID shall be capable of multi-beaming, providing simultaneous access to large, possibly non-contiguous and distantly separated parts of the sky.*

Requirement AMSC_R1004a is based on the recommendation in RD-3 for transients science. A very large FoV significantly increases the detection rate of transients (e.g. Macquart 2014; Macquart et al. 2015), particularly for the rarest, brightest events. Moreover, a FoV of this size would allow the sample of 10 000 FRBs needed for cosmological studies to be found within 2-3 years of on-sky telescope time, based on the current best-estimate rates (e.g. Keane & Petroff 2015).Multi-beaming significantly enhances the flexibility of SKA-AAMID. Specific science examples include bulk timing of a suite of pulsars spread across the sky, and wide-field searches for transients and variables. Being able to carry out independent observing programmes in parallel is also highly desirable.

**AMSC_R1005**:

*SKA-AAMID shall provide access to all local azimuthal angles, 0° – 360°, and all local elevation angles, 45° – 90°, to ensure a minimum total accessible sky coverage of approximately 25 000 deg$^2$. There is a desire to observe at elevation angles 30° – 45° to extend the sky coverage to approximately 30 000 deg$^2$ for at least a subset of observing frequencies .*

This requirement is necessary for all-sky science cases such as cosmology, as well as searches for transients and pulsars. The Large and Small Magellanic Clouds will also be accessible (Declinations of -69.8° and -72.8°, respectively), allowing important studies on topics such as gas accretion and star formation. The suggested extension of the scan angle to elevations as low as 30° would ensure that the entire Southern hemisphere is accessible to the telescope; for example, observing the South Celestial Pole would be advantageous for (semi-)continuous searches for transients. Furthermore, the full ecliptic would be accessible for Solar imaging, and there would also be an increased sky overlap with a number of next-generation facilities operating in the Northern hemisphere.

**AMSC_R1006**:

*SKA-AAMID shall provide nominal sensitivity at all local azimuthal angles, 0 – 360°. Sensitivity will be optimized at zenith (Declination ≈ -30°). Sensitivity degradation away from zenith will remain within 5 per cent of the theoretical performance loss of the system due to the changing geometric foreshortening of the aperture.*

This requirement provides an initial performance benchmark to ensure reasonable sensitivity levels away from zenith, in turn allowing both all-sky surveys and individual deep pointings to be carried out in an acceptable period of time. Optimizing the sensitivity at zenith is advantageous because a significant fraction of the Galaxy passes close to or directly overhead; in addition, there is also good coverage of high Galactic latitudes, including the South Galactic Pole, which is an excellent candidate deep field location for extragalactic projects.

**AMSC_R1007**:

*SKA-AAMID shall support simultaneous imaging and beamformed observations.*

This requirement is mainly linked to the transients science case. Depending on the sub-class of transient, and the degree of scatter broadening, having both interferometric and beamformed data offers the most complete picture of an outburst (e.g. Hassall et al. 2013). Another important example is detecting an FRB in beamformed data, and using interferometric data to localize it (e.g. Macquart et al. 2015). Furthermore, commensality is absolutely key to transient science, and so this observing mode should be used for all observations, where possible.

**AMSC_R1008:**

This requirement was deleted based on conclusions of the System Requirements Review (SRR), 2016 July 13-14.

**AMSC_R1009:**

*SKA-AAMID shall be capable of making continuum images, with any specific integration time in any specific direction within its sky coverage, with an effective noise level that is within a factor of 2 of the thermal noise level expected theoretically based on telescope sensitivity, bandwidth, integration time and the weighting scheme used in the imaging process.*

This requirement is adapted from AD-13, and allows a range of science goals to be met, particularly for deep imaging.

**AMSC_R1010**:

*SKA-AAMID shall provide an absolute photometric accuracy ≤ 1–2 per cent, and an internal (relative) systematic photometric accuracy ≤ 1 per cent.*

This requirement is adapted from AD-13. Following AD-13, photometric accuracy is defined as the fractional error, $\Delta S/S$, relative to an adopted celestial flux standard for the integrated brightness of compact objects (smaller than the PSF FWHM). The absolute requirement is based on the latest state-of-the-art flux density scales (e.g. Perley & Butler 2013). The internal consistency of flux densities, over both time and frequency, across a range of observing runs, is vital for many different science applications, particularly studies of variability.

**AMSC_R1011**:

*SKA-AAMID shall use an absolute astrometric reference frame that is accurate to ≤ 10 μas. The internal (relative) systematic astrometric accuracy shall be ≤ 1 per cent.*

This requirement is adapted from AD-13. The absolute requirement is based on Fomalont & Reid (2004); also see Paragi et al. (2015). Following AD-13, astrometric accuracy is defined as the fractional error in the position, $\Delta \rho / \Delta \theta$, for the centroid of compact objects (smaller in size than the PSF FWHM, $\Delta \theta$) relative to the adopted celestial reference frame. The internal consistency of astrometric positions, over both time and frequency, across a range of observing runs, is vital for many different science applications, for example proper motion studies.

**AMSC_R1012**:

This requirement was deleted based on conclusions of the System Requirements Review (SRR), 2016 July 13-14.

**AMSC_R1013**:

*SKA-AAMID shall provide a frequency accuracy ≤ $10^{-11}$ over a 10 year interval.*

This requirement is taken from AD-13, and mainly relates to the real-time cosmology experiment outlined in Kloeckner et al. (2015). Following AD-13, frequency accuracy is defined as the fractional error in the frequency, $\Delta \nu / \nu$, relative to an adopted frequency standard over a specified time interval.

**AMSC_R1014**:

*SKA-AAMID shall provide timing accuracy ≤ 10 ns over a 10 year interval.*

This requirement is taken from AD-13, and is particularly relevant for high-precision pulsar timing (e.g. RD-3; Janssen et al. 2015). Following AD-13, timing accuracy is defined as the **RMS** time error, $\Delta \tau$, relative to an adopted time standard over an indicated time interval.

**AMSC_R1015**:

*SKA-AAMID shall have sufficient uv coverage and surface brightness sensitivity to enable transformational science on a range of angular scales: from several tens of $deg^2$, to ≤ 0.5 arcsec in angular resolution.*

Proposed experiments for SKA-AAMID range from IM on very short intra-station baselines, to sub-arcsec angular resolution continuum observations. The collecting area of the telescope must be distributed sufficiently so as to ensure that the widest range of primary scientific objectives can be met. As an example, a collecting area distribution of 50 per cent within 5 km would be ideal for pulsar and fast transient studies (e.g. Smits et al. 2008,2009).

## 5.2. Galaxy evolution and cosmology probed by neutral hydrogen

**AMSC_R2001**:

*SKA-AAMID shall achieve a Stokes-I sensitivity ≤ 5 µJy beam$^{-1}$ RMS from 450-1425 MHz, over a bandwidth of 10 kHz, with 2-15 arcsec angular resolution, over at least 25 000 deg$^2$ and up to a maximum of 30 000 deg$^2$, within two years on-sky integration.*

This requirement will allow one of the original primary science visions of the SKA to be achieved: a billion galaxy HI survey out to a redshift of 2 and beyond. Such a survey will enable unprecedented studies of galaxy evolution, as well as world-class precision cosmology, such as through the detection of baryon acoustic oscillations. This requirement is based on Santos et al. (2015b) and RD-3. Following Santos et al. (2015b), a minimum detection is assumed to be at the 10σ level.

**AMSC_R2002**:

*SKA-AAMID shall achieve a Stokes-I sensitivity ≤ 1.5-5 µJy beam$^{-1}$ RMS from 450 – 650 MHz, over a bandwidth of 10 kHz, with 2 arcsec angular resolution, over a selection of fields with total area 30-50 deg$^2$, within 1000 hours on-sky integration per field.*

This requirement will enable a fiducial HI mass of $M^* = 5 \times 10^9$ solar masses (or lower) to be reached in deep fields at high redshift (RD-3), assuming averaging is carried out over a feature that would have a velocity width of 50 km s$^{-1}$ in the local Universe, as well as the velocity width - redshift scaling assumptions used by Abdalla & Rawlings (2005). For consistency with AMSC_R2001, a minimum detection is assumed to be at the 10σ level, and the lower and upper flux densities stated in the requirement correspond to the required values at 450 and 650 MHz, respectively.

**AMSC_R2003**:

*SKA-AAMID shall achieve a Stokes-I sensitivity ≤ 20 µJy beam$^{-1}$ RMS from 450 – 1425 MHz, over a channel width of 2 kHz, with an angular resolution ≤ 1 arcsec, over at least 10 000 deg$^2$, within two years on-sky integration.*

This requirement will allow a wide field, blind HI absorption survey to be performed to optical depths of e.g. 0.01/0.001 (10 mJy / 100 mJy background continuum sources; assuming 5σ detections), and to redshifts beyond 2 (RD-3; Morganti et al. 2015).

**AMSC_R2004**:

*SKA-AAMID shall achieve a Stokes-I sensitivity ≤ 15 µJy beam$^{-1}$ RMS from 1400 – 1425 MHz, over a bandwidth of 5 kHz (velocity resolution Δv ≈ 1 km s$^{-1}$), with 1-10 arcsec angular resolution, over an area of approximately 20 deg$^2$, within 10 hours on-sky integration.*

This requirement will allow highly-detailed studies of the interstellar medium in nearby galaxies, including at 50 pc resolution for galaxies within 10 Mpc, as well as their immediate ambient environments, to column density limits of ≤ $10^{17-19}$ cm$^{-2}$ (e.g. de Blok et al. 2015; column density calculation using equation 3 in Fernández et al. 2013). Note that the requirement refers to an observation of each individual galaxy, and assumes a minimum 5σ detection level.

**AMSC_R2005**:

*SKA-AAMID shall achieve a Stokes-I sensitivity ≤ 1.7 µJy beam$^{-1}$ RMS from 1400 – 1425 MHz, over a bandwidth of 50 kHz (Δv ≈ 10 km s$^{-1}$), with 10 arcsec angular resolution, over at least 25 000 deg$^2$ and up to a maximum of 30 000 deg$^2$, within two years on-sky integration.*

This requirement will allow detections of diffuse gas filaments in the nearby cosmic web with a spatial resolution of a few kpc and better, at column densities ≤ 10$^{18}$ cm$^{-2}$ (e.g Popping et al. 2015; column density calculation using equation 3 in Fernández et al. 2013). A minimum detection level of 5σ is assumed.

**AMSC_R2006**:

*Between 450 – 1425 MHz, SKA-AAMID shall have a velocity resolution ≤ 2 km s$^{-1}$.*

This is required for sufficient velocity resolution for the full range of extragalactic HI science cases (e.g. RD-3). Note in particular that the proposed real-time cosmology experiment outlined in Kloeckner et al. (2015) would require velocity resolution as fine as 2.9 cm s$^{-1}$.

**AMSC_R2007**:

*SKA-AAMID shall achieve a spectral dynamic range ≥ 60 dB from 450 – 1425 MHz, over a bandwidth ≤ 10 kHz.*

This requirement is necessary for deep HI detections (e.g. RD-3). Following AD-13, spectral dynamic range is defined as the ratio of peak brightness to the residual instrumental spectral response, I/ΔI, at the source location over a specified spectral interval. The minimum bandwidth requirement follows from AMSC_R2001 and AMSC_R2002.

**AMSC_R2008**:

*SKA-AAMID shall achieve a bandpass stability of 1 in 10$^6$ for integration times of up to 1000 hr per field.*

This requirement is necessary to unambiguously detect HI emission/absorption at the tens of µJy level. Note that we assume the same definition of bandpass stability as in AD-13: the residual fractional error in the brightness, ΔI/I, as function of frequency for a specified frequency interval over a specified time.

## 5.3. Galactic HI

**AMSC_R3001**:

*SKA-AAMID shall achieve a Stokes-I sensitivity ≤ 20 µJy beam$^{-1}$ RMS from 1415 – 1425 MHz, over a bandwidth of 1.4 kHz (Δv ≈ 0.3 km s$^{-1}$), with 5 arcsec angular resolution, over at least 25 000 deg$^2$ and up to a maximum of 30 000 deg$^2$, within six months on-sky integration.*

This requirement will allow the all-sky Galactic HI emission survey for mapping the structure of turbulence in the interstellar medium down to length scales on the order of the size of the Solar system (McClure-Griffiths et al. 2015).

**AMSC_R3002**:

*Between 1415 and 1425 MHz, SKA-AAMID shall have a velocity resolution of 0.3 km s$^{-1}$.*

This is required to resolve narrow HI absorption lines in the Galactic cold neutral medium which have line widths on the order of 2.5 km s$^{-1}$ (McClure-Griffiths et al. 2015).

## 5.4. The transient radio sky

**AMSC_R4001**:

*SKA-AAMID shall achieve a Stokes-I sensitivity ≤ 1 mJy beam$^{-1}$ RMS from 450 – 1450 MHz, over a bandwidth of 100 MHz, with ≤ 2 arcsec angular resolution, over the full instantaneous processed FoV, for an integration time of 1 ms.*

This level of sensitivity for coherent, fast transients will allow for a far more complete understanding of transient parameter space (e.g. Cordes et al 2004; RD-3). Note that the total instantaneous bandwidth may be much larger than 100 MHz, but the requirement then ensures that a radio spectrum can be determined. De-dispersion of the signal is assumed. High angular resolution is needed for accurate transient localization.

**AMSC_R4002**:

*SKA-AAMID shall achieve a Stokes-I sensitivity ≤ 5 µJy beam$^{-1}$ RMS from 450 – 1450 MHz, over a bandwidth of 100 MHz, with ≤ 2 arcsec angular resolution, over at least 25 000 deg$^2$ and up to a maximum of 30 000 deg$^2$, within 24 hours on-sky integration.*

Deep, high-cadence, all-sky surveys will allow for a far more complete understanding of transient parameter space (e.g. Cordes et al 2004; RD-3). Note that the total instantaneous bandwidth may be much larger than 100 MHz, but the requirement then ensures that a radio spectrum can be determined. High angular resolution is needed for accurate transient localization.

**AMSC_R4003**:

*The SKA-AAMID data streams (imaging and beamformed) shall be automatically searched for astrophysical transients, with unambiguous detections being reported with a latency of less than 30 seconds at most, and preferably less than 10 seconds.*

Low-latency capabilities are mandatory for optimizing this area of science (e.g. Fender et al. 2015a; Chu et al. 2016). Rapid dissemination of alerts maximises the global scientific return via multi-wavelength follow up.

**AMSC_R4004**:

*SKA-AAMID shall be able to respond to both human- and robotically-generated alerts on a time-scale of 5 seconds at most, and preferably less than 1 second, and begin beamforming in the appropriate direction(s) on this time-scale.*

This requirement is essential for high-impact science (e.g. Fender et al. 2015a). SKA-AAMID will then be able to fully participate in multi-facility campaigns to understand the most exciting, novel astrophysics from outbursts, which is often at most very shortly after the event itself.

**AMSC_R4005**:

*SKA-AAMID shall be able to store raw voltage data over the full optical FoV and for a (possibly non-contiguous) bandwidth ≥ 250 MHz, for a period ≥ 120s.*

This will ensure proper localization and follow-up of fast transients, including the rise phase of a given transient event. As an example, current FRB searches at Parkes make use of a 120 s ring buffer (e.g. Petroff et al. 2015). The maximum detectable DM would be approximately 6500 pc cm$^{-3}$, which would generally be comfortably adequate for most purposes (e.g. Cordes & Lazio 2002; Petroff et al. 2016).

**AMSC_R4006**:

*The temporal resolution of SKA-AAMID beamformed data shall be on a time-scale as short as 100 nanoseconds. It is also desirable to have a temporal resolution as short as nanoseconds.*

This will allow the fullest sampling of transients parameter space (e.g. Cordes et al. 2004; RD-3), with an ultra-short time-scale of nanoseconds corresponding to very high brightness temperature events such as giant pulses from the Crab Pulsar (e.g. Staelin & Sutton 1970).

**AMSC_R4007**:

*SKA-AAMID shall have sufficient instantaneous uv coverage as to allow the localization of fast transients in imaging data to a precision ≤ 0.1 arcsec for signal-to-noise ratios ≥ 5, and an integration time of 50 ms. It is also desirable to be able to make similar images for integration times < 50 ms.*

This requirement is necessary for fast transient science, such as FRB studies (e.g. Macquart et al. 2015). Following Fomalont (1999), being able to synthesize a 1 arcsec beam on these time-scales would allow a formal localization as precise as 0.1 arcsec for a 5σ detection.

## 5.5. Strong-field tests of gravity with pulsars

**AMSC_R5001**:

*SKA-AAMID shall achieve a Stokes-I 1400 MHz limiting luminosity of 0.1 mJy kpc$^2$ (or equivalent over the frequency range 450 – 1450 MHz) at distances ≥ 5 kpc, over a bandwidth ≤ 100 MHz, over at least 25 000 deg$^2$ and up to a maximum of 30 000 deg$^2$, within two years on-sky integration.*

This requirement will allow SKA-AAMID to detect the vast majority of the visible pulsar population of the Galaxy (but excluding the population at the Galactic Centre). The luminosity requirement is based on the work presented in Lorimer et al. (2006) and Smits et al. (2009), as well as RD-3. At a distance of 5 kpc, this corresponds to a detectable 1400 MHz flux density requirement of 4 μJy (0.8 μJy beam$^{-1}$ RMS for a 5σ detection) . Alternatively, assuming 10 kpc (e.g. as in RD-3) gives a 1400 MHz flux density requirement of 1 μJy (0.2 μJy beam$^{-1}$ RMS). Pulsar surveys will typically have several hundred MHz bandwidth or more that could be used to meet the sensitivity requirements (e.g. RD-3; Smits et al. 2009), but ideally detections would be made using e.g. no more than 100 MHz bandwidth, to allow a spectral index to be determined. Note that higher frequencies are more appropriate for lower Galactic latitudes, where (i) the dispersion measure will generally be larger; (ii) the effects of scattering can be significant at lower frequencies; and (iii) the sky temperature is higher, particularly at lower frequencies. Assuming an average pulsar spectral index of -1.4 (Bates et al. 2013), this would result in sensitivity requirements that are generally less stringent at lower frequencies.

**AMSC_R5002**:

*Using the full instantaneous bandwidth from 450 – 1450 MHz, SKA-AAMID shall carry out initial bulk timing of all pulsars discovered by the telescope, taking no more than approximately 250 h on-sky integration over a six month period to achieve a signal-to-noise ratio ≥ 10 per pulsar over a total sky area of about 15 000 $deg^2$.*

The observing strategy necessary for such a programme is discussed in Smits et al. (2009), as well as Lorimer et al. (2006). For example, individual pointings will typically last several minutes, reaching 1400 MHz luminosities of about 0.1 – 0.5 mJy $kpc^2$ (i.e. approximately 0.3 µJy $beam^{-1}$ RMS at 10 kpc).

**AMSC_R5003**:

*SKA-AAMID shall have the capability to track dispersion measure variations of magnitude ≤ $10^{-4}$ pc $cm^{-3}$, associated with very-high-precision pulsar timing experiments, using the full bandwidth range 450 – 1450 MHz,*

This requirement is based on the recommendations in RD-3 for the pulsar timing array (also see e.g. Janssen et al. 2015). Examples of dispersion measure variations for the current generation of pulsar timing array candidates can be found in e.g. Keith et al. (2013); also see e.g. Lee et al. (2014).

**AMSC_R5004**:

*SKA-AAMID shall achieve a polarization purity ≤ -40 dB, on axis.*

This requirement follows the recommendation in RD-3, to ensure sufficiently precise pulsar timing. Polarization purity is defined using AD-13 as a starting point: the ratio of the residual instrumental polarized response peak to Stokes I brightness to, ΔP/I, at the source location. Here, the specified value applies to the target pointing directions at the centres of individual beams.

**AMSC_R5005**:

*SKA-AAMID beamformed modes shall have a frequency resolution ≤ 10 kHz per channel.*

This requirement follows the recommendation in RD-3, to minimize dispersion smearing.

**AMSC_R5006**:

*SKA-AAMID shall have beamformed observational modes with a temporal resolution between 100 ns – 50 µs. It is desirable to have temporal resolution better than 100 ns.*

The stated minimum temporal resolution follows the recommendation in RD-3 for a general pulsar survey (using a dense array core), and would also potentially allow for sub-millisecond pulsars to be discovered, if they indeed exist. Also, following RD-3, 100 ns temporal resolution would allow the pulse profiles of millisecond pulsars to be resolved by at least a factor of a few hundred, allowing the best candidates for the pulsar timing array to be identified.

## 5.6. Galaxy evolution and cosmology probed by radio continuum

**AMSC_R6001**:

*SKA-AAMID shall achieve a Stokes-I sensitivity ≤ 100 nJy beam$^{-1}$ RMS from 450 – 1450 MHz, over a bandwidth ≤ 300 MHz, with angular resolution ≤ 0.5 arcsec at all frequencies and as good as 0.2 arcsec above 1000 MHz (with the possibility of extending this resolution to as low as 600 MHz), over at least 25 000 deg$^2$ and up to a maximum of 30 000 deg$^2$, within 6 months on-sky integration.*

This requirement allows a wide field, high-resolution radio continuum survey to be carried out, in which over three billion galaxies will be detected, allowing unprecedented studies of star formation across cosmic time, as well as cosmological constraints to be derived from both weak and strong lensing (e.g. Jarvis et al. 2015a,b; Brown et al. 2015; Patel et al. 2015; McKean et al. 2015; Norris et al. 2015). A minimum detection level of 5σ is assumed. Depending on how the survey strategy is refined, a canonical spectral index of -0.7 would result in sensitivity requirements that are less stringent by a factor of approximately 2 at 600 MHz (the lowest frequency for the survey assumed in Prandoni & Seymour 2015) than at 1450 MHz.

**AMSC_R6002**:

*SKA-AAMID shall achieve a Stokes-I sensitivity ≤ 5 nJy beam$^{-1}$ RMS from 450 – 1450 MHz, over a bandwidth ≤ 300 MHz, with angular resolution ≤ 0.5 arcsec at all frequencies and as good as 0.2 arcsec above 1000 MHz (with the possibility of extending this resolution to as low as 600 MHz), over a selection of fields with total area 30 – 50 deg$^2$, within 1000 hours on-sky integration per field.*

This requirement will enable highly-detailed studies of star formation (e.g. to a precision of 25 solar masses per year at z ∼ 7) and AGN activity in selected ultra-deep fields (e.g. RD-3, Jarvis et al. 2015b). A minimum detection level of 5σ is assumed. Depending on how the survey strategy is refined, a canonical spectral index of -0.7 would result in sensitivity requirements that are less stringent by a factor of approximately 2 at 600 MHz (the lowest frequency for the observations assumed in Prandoni & Seymour 2015) than at 1450 MHz. The suggested angular resolution of 30 mas by Jarvis et al. (2015b) could be achieved by a hybrid combination of AAMID stations and dishes (Section 6).

**AMSC_R6003**:

*SKA-AAMID shall achieve an imaging dynamic range ≥ 70 dB from 450 – 1450 MHz, over a bandwidth ≤ 300 MHz.*

This requirement is necessary for ultra-deep radio continuum imaging. Following AD-13, imaging dynamic range is defined as the ratio of peak to RMS brightness, I/ΔI, where the RMS fluctuation level is measured within the entire station beam(s), but may exclude an area of 5x5 PSF FWHM centred on the brightness peak.

**AMSC_R6004**:

*For radio continuum surveys and deep imaging conducted over the frequency range 600 – 1450 MHz, where the theoretical noise level is ≤ 100 nJy beam$^{-1}$, the measured RMS noise level in SKA-AAMID images shall not be limited by source confusion.*

The confusion estimates presented in Condon et al. (2012) imply that an angular resolution better than a few arcsec is generally needed, which is consistent with requirements AMSC_R6001 and AMSC_R6002. However, more specific requirements may result from continuum observations carried out with the SKA pathfinders, as well as SKA1.

## 5.7. The origin and evolution of cosmic magnetism

**AMSC_R7001**:

*SKA-AAMID shall achieve a Stokes Q and U sensitivity ≤ 100 nJy beam$^{-1}$ RMS from 450 – 1450 MHz, with (a possibly non-contiguous) instantaneous bandwidth ≥ 500 MHz, with 2-3 arcsec resolution, over at least 25 000 deg$^2$ and up to a maximum of 30 000 deg$^2$, within six months on-sky integration.*

This requirement will enable an all-sky grid of Faraday rotation measures to be constructed, with a density of several thousand per square degree, an order of magnitude denser than in SKA1. Applications include measurement of the magnetic fields in a significant number of intervening galaxies out to high redshifts, as well as highly detailed measurements of the structure of the magnetic field in the Milky Way (e.g. RD-3; Johnston-Hollitt et al. 2015; Gaensler et al. 2015; Haverkorn et al. 2015). A minimum detection level of 10σ is assumed for sufficiently accurate RM synthesis (e.g. RD-3); note that the full 1000 MHz bandwidth would yield the optimal point spread function in Faraday depth space.

**AMSC_R7002**:

*SKA-AAMID shall achieve a Stokes Q and U sensitivity ≤ 10 nJy beam$^{-1}$ RMS from 450 – 1450 MHz, with (a possibly non-contiguous) instantaneous bandwidth ≥ 500 MHz, with ≤ 3 arcsec resolution, over a selection of fields with total area 30 – 50 deg$^2$, within 1000 hours on-sky integration per field.*

This requirement will allow the origin and evolution of cosmic magnetism to be explored at very faint flux density levels in selected ultra-deep fields (RD-3; Johnston-Hollitt et al. 2015; Gaensler et al. 2015; Taylor et al. 2015). A minimum detection level of 10σ is assumed for sufficiently accurate RM synthesis (e.g. RD-3); note that the full 1000 MHz bandwidth would yield the optimal point spread function in Faraday depth space.

**AMSC_R7003**:

*SKA-AAMID shall achieve a Stokes Q and U sensitivity ≤ 20 nJy beam$^{-1}$ RMS from 450 – 1450 MHz, with (a possibly non-contiguous) instantaneous bandwidth ≥ 500 MHz, with 0.5-1.5 arcsec resolution, over an area of approximately 20 deg$^2$, within 100 hours on-sky integration.*

This requirement is necessary for significant improvements in magnetic field tomography of nearby galaxies (e.g. Heald et al. 2015), compared with SKA1. In particular, the evolution of the three-dimensional magnetic field structure in galaxies will be investigated at a spatial resolution of as good as 10 pc for nearby galaxies, 100 pc for all galaxies within 50 Mpc, and on scales of 1 kpc out to a redshift of about 0.1. Note that the requirement refers to an observation of each individual galaxy. The final resolution chosen across the frequency range will depend on the required sensitivity to extended emission (see discussion in Heald et al. 2015).

**AMSC_R7004**:

*SKA-AAMID shall achieve a polarization purity of at least -25 dB, and preferably as good as -30 dB, over the full processed FoV.*

This requirement follows the recommendations in RD-3 and Taylor et al. (2015), to ensure sufficiently precise studies of the faint polarized radio sky. Polarization purity is defined as in AMSC_R7004.

**AMSC_R7005**:

*SKA-AAMID shall be able to routinely achieve an observed-frame Faraday depth precision ≤ 1 rad m$^{-2}$, and to be able to detect observed-frame Faraday depths up to a maximum amplitude of 10 000 rad m$^{-2}$.*

This requirement will allow SKA-AAMID to be sensitive to a wide range of magnetized phenomena. The first part can be achieved by utilizing the full frequency coverage, 450 - 1450 MHz; for example, the formal RM error on a 10 sigma detection would be 0.43 rad m$^{-2}$ For the second part of the requirement, the channel width at 450 MHz would need to be 90 kHz, using the relevant formula in Brentjens and de Bruyn (2005); at 1450 MHz, a channel width of 3 MHz would suffice.

## 5.8. Widefield SETI

**AMSC_R8001**:

*SKA-AAMID shall achieve a sensitivity ≤ 190 mJy beam$^{-1}$ RMS per polarization from 450 – 1450 MHz, over a bandwidth ≤ 0.01 Hz, within 10 min on-sky integration.*

This requirement is for narrowband SETI, in particular airport radar-like signals from stars within 60 pc (Siemion et al. 2015). Following Siemion et al. 2015, a minimum detection is assumed to be at the 12σ level.

## 6. Recommendation for SKA-AAMID parameters

Using the science requirements from Section 4, key parameters leading to an overall recommended design are summarised in Table 1.

Table 1: Key SKA-AAMID parameters that define the performance envelope of the system.

| Parameter | Value (required) | Value (goal) | Science experiment(s) that drive(s) the requirement |
|---|---|---|---|
| Lowest frequency (MHz) | 450 | 400 | High-z HI |
| Highest frequency (MHz) | 1450 | 1500 | Range of experiments |
| Instantaneous sensitivity at zenith $A_{eff}/T_{sys}$ (m$^2$/K) | 10 000 | 15 000 | HI / continuum deep fields |
| Optical FoV (deg$^2$) | 200 (at 1000 MHz) | 250 (at 1000 MHz) | Transients |
| Processed FoV (deg$^2$) and instantaneous bandwidth (MHz) [product; can be traded off against each other] | ≥100 (at 1000 MHz; 500 MHz bandwidth per beam) | ≥100 (at 1000 MHz; 1000 MHz bandwidth per beam) | Range of experiments |
| Total instantaneous bandwidth (MHz) | 1000 (per processed beam) | 1000 (per processed beam) | Range of experiments |
| Minimum baseline (m) | 20 | 1 | Correlations within the core station for HI intensity mapping; Galactic HI |
| Maximum angular resolution (arcsec) | 0.5 | 0.2 | Continuum surveys, including weak/strong lensing |
| Scan angle (deg; from zenith) | ±45 | ±60 | Cosmology; transients; pulsar searches; Galactic HI (Small and Large Magellanic Clouds) |
| Overall sky coverage (deg$^2$) | 25 000 deg$^2$ | 30 000 deg$^2$ | Cosmology; transients; pulsar searches |
| Native temporal resolution (s) | 0.15 (interferometric; standard imaging mode) | 0.05 (interferometric; fast imaging mode) | Range of experiments; possible fast imaging mode for transients |
| | $10^{-7} - 10^{-5}$ (beamformed) | $10^{-9} - 10^{-5}$ (beamformed) | Pulsars; transients |
| Native frequency resolution (kHz) | 1.4 (interferometric) | 1.4 (interferometric) | Galactic HI |
| | ≤10 (beamformed) | ≤$10^{-5}$ (beamformed; correlator in zoom mode with narrow bandwidth) | Pulsars; transients. The goal of ≤$10^{-5}$ kHz (≤ 0.01 Hz) is for real-time measurement of space-time expansion, and SETI. |
| Frequency agility | Option for non-contiguous bandwidth | Option for non-contiguous bandwidth | Transients; pulsars; cosmic magnetism; commensal observing |

| Imaging dynamic range (dB) | 70 | 75 | Continuum deep fields |
|---|---|---|---|
| Spectral dynamic range (dB) | 60 | 65 | HI BAO and galaxy evolution |
| Polarization capability | Full Stokes | Full Stokes | All experiments |
| Polarization purity (dB) | -25 (off axis) -40 (on axis) | -30 (off axis) -40 (on axis) | Pulsar timing ; cosmic magnetism |
| Collecting area distribution | 50% collecting area within 5km | 50% collecting area within 5km | Pulsars; transients; HI intensity mapping |
| Array reconfiguration time (s) | <5 | <1 | Transients |

# 7. Progressive science operations with the SKA-AAMID telescope

**Phased approach to AAMID implementation: initial operations with SKA1-MID**

Several key science projects are optimally addressed with a large fraction of the collecting area in the core. These include pulsar searches, fast transient searches, the Galactic HI survey, and the Intensity Mapping experiment for the detection of the BAO signature. SKA-AAMID is perfectly suited to these surveys, providing an extremely large FoV, and the possibility to correlate very short baselines for sensitivity to extended emission.

The large FoV necessitates a significant digital processing capability, especially for surveys at high angular resolution. The combination of large FoV and high angular resolution results in a very large number of beams on the sky. In order to remain within processing limitations, only a subset of the FoV can be processed with high angular resolution. As a result, much of the RF capability of SKA-AAMID is not exploited when doing surveys with the longest baselines. It is natural to have an early implementation SKA architecture that is composed of AAMID stations in the core, and with the SKA1-MID dishes in the outer stations. However, future upgrades to processing capacity will eventually catch up with the RF capability, delivering the full potential of SKA-AAMID as the ultimate survey machine.

A hybrid architecture uses the advantages of both frontend technologies. The AAMID core provides a very large FoV, while the outer dishes increase the angular resolution for mapping within a subset of the AAMID FoV. The dishes must point to lower elevations to match the core pointing, but this is not an issue in general as they will be able to slew well below, for example, a 45 degree elevation angle (e.g. RD-3).

A further consideration is the exploitation of existing instruments. When SKA-AAMID is being installed, there will already be SKA1-MID operating. In the frequency overlap region, the first AAMID station could begin operations immediately as part of SKA1, providing a large core to the existing infrastructure. This will significantly augment the capability of SKA1-MID by providing increased sensitivity to the overall instrument, and, by immediately beginning a transient survey. SKA1 with AAMID will be a powerful instrument for transient detections and immediate follow-up observations.

In the nearer term, an AAMID technology demonstrator can be installed and running together with MeerKAT. This combination would already be the most powerful transient survey instrument in operation.